\documentclass{aastex63}

\received{April 3, 2020}
\revised{May 22, 2020}
\accepted{May 28, 2020}

\submitjournal{ApJL}

\shorttitle{Evidence for a Past Martian Ring}
\shortauthors{{\'C}uk et al.}

\begin{document}

\title{Evidence for a Past Martian Ring from the Orbital Inclination of Deimos}

\correspondingauthor{Matija {\'C}uk}
\email{mcuk@seti.org}

\author{Matija {\'C}uk}
\affil{SETI Institute \\
189 North Bernardo Ave, Suite 200 \\
Mountain View, CA 94043, USA}

\author{David A. Minton}
\affiliation{Department of Earth, Atmospheric and Planetary Sciences \\
Purdue University  \\
550 Stadium Mall Drive\\ 
West Lafayette, IN 47907, USA}

\author{Jennifer L.~L. Pouplin}
\affiliation{Department of Earth, Atmospheric and Planetary Sciences \\
Purdue University  \\
550 Stadium Mall Drive\\ 
West Lafayette, IN 47907, USA}

\author{Carlisle Wishard}
\affiliation{Department of Earth, Atmospheric and Planetary Sciences \\
Purdue University  \\
550 Stadium Mall Drive\\ 
West Lafayette, IN 47907, USA}


\begin{abstract}
We numerically explore the possibility that the large orbital inclination of the martian satellite Deimos originated in an orbital resonance with an ancient inner satellite of Mars more massive than Phobos. We find that Deimos's inclination can be reliably generated by outward evolution of a martian satellite that is about 20 times more massive than Phobos through the 3:1 mean-motion resonance with Deimos at 3.3 Mars radii. This outward migration, in the opposite direction from tidal evolution within the synchronous radius, requires interaction with a past massive ring of Mars. Our results therefore strongly support the cyclic martian ring-satellite hypothesis of \citet{hes17}. Our findings, combined with the model of \citet{hes17}, suggest that the age of the surface of Deimos is about 3.5-4~Gyr, and require Phobos to be significantly younger.
\end{abstract}


\section{Introduction} \label{sec:intro}
Two small satellites of Mars, Phobos and Deimos, are thought to have formed from debris ejected by a giant impact onto early Mars \citep{cra11, ros12, cit15, ros16, hes17, hyo17, can18}. The small masses of the satellites, compared to expectations from giant impact simulations, has lead to proposals that the original satellite system was more massive, with most of its mass falling onto Mars over several Gyr following the impact \citep[see ][for an alternate view]{can18}. The early massive satellite hypothesis offers at least two competing explanations of Phobos's short \citep[40~Myr;][]{lai07} lifetime against inward tidal evolution. \citet{cra11} and \citet{ros16} suggest that several satellites formed from the impact-generated debris disk and most have since tidally decayed onto Mars, with Phobos and Deimos being the last survivors due to their formation just interior and exterior to the martian synchronous orbit, respectively. Alternatively, \citet{hes17} propose that Phobos is only the latest product of a repeating ring-satellite cycle at Mars, with each successive inner satellite being less massive than the preceding one. In the ring-satellite cycle model, satellites form from the outer edge of the ring, and then migrate outward through gravitational interaction with the ring. The ring loses mass to the planet at its inner edge, and once the ring is sufficiently depleted the satellite migrates inward due to tides. The inward migration of the satellite continues until tidal stresses disrupt it to form a new ring \citep[cf. ][]{bla15, hes19}, which viscously spreads, thus re-starting the cycle.   

One difficulty for the giant impact scenario is the relatively large distance at which Deimos orbits Mars, beyond the martian synchronous orbit. In the ring-satellite cycle model of \citet{hes17}, only satellites interior to the synchronous orbit would cycle, and therefore Deimos would have to have formed from the original impact-generated circumplanetary disk, as in \citet{cra11}. Alternatively, \citet{ros16} have suggested that Deimos (or its precursors) may have been ``pushed out" through a resonance with an ancient massive inner satellite. While attractive, this scenario may have difficulty in producing Deimos on its present orbit, as we would expect the orbit of a resonantly-pushed satellite to be quite excited \citep[e.g.][]{mal93}, and tidal dissipation within Deimos may be too inefficient to damp its eccentricity over the age of the solar system \citep{sze83}. While early work on Deimos's cratering record estimated that its surface dates back to the end of the Late Heavy Bombardment \citep{tho80}, that work did not consider planetocentric and, especially, sesquinary impacts. \citet{nay16} show that the sesquinary impacts on Deimos are an important source of cratering, making determination of its surface age very difficult. This would be particularly true if Deimos had a more eccentric or inclined orbit in the past, which would increase relative energy of sesquinary impactors which have similar eccentricities and inclinations as Deimos, but whose orbits have precessed out of alignment with it. Therefore, if the orbit of Deimos was highly eccentric and inclined in the past, it could have suffered runaway sesquinary bombardment, which in turn could have damped Deimos's $e$ and $i$, potentially reconciling an early period of strong dynamical excitation with its current near-circular/equatorial orbit.

The present-day orbit of Deimos can be used to constrain the martian satellites' dynamic past. The current mean inclination of Deimos relative to the Laplace plane is $i_f=1.8^{\circ}$, while its mean eccentricity is only $e = 2.7 \times 10^{-4}$ \citep{jac14}. \citet{yod82} found that the inward orbital migration of Phobos through the satellites' mutual 2:1 mean-motion resonance (MMR) excites the eccentricity of Deimos to $e =0.002$, an order of magnitude higher than it is now. \citet{yod82} speculated that higher-than-expected tidal damping of eccentricity by Deimos (possibly due to resonantly excited rotational librations) could explain this discrepancy, but available data on the moments of inertia of Deimos do not support exceptionally large librational response \citep{tho93}. Recently, \citet{qui20} proposed that Deimos could have experienced enhanced dissipation due to chaotic rotation \citep[c.f.][]{wis87} down to very low eccentricities. However, we think that the results of \citep{qui20} are an artifact of their accelerated simulations, in which the eccentricity damping timescale was artificially made shorter than the spin-down and wobble damping timescales. The problem of Deimos's eccentricity arises for any scenario in which Phobos starts its inward evolution beyond the location of 2:1 resonance at 4.3 $R_M$. In such ``ancient'' Phobos model, including those of \citet{cra11}, \citet{ros16}, and \citet{can18}, Phobos  should have crossed the 2:1 resonance with Deimos and excited its orbit about 2~Gyr ago \citep{yod82}. On the other hand, the cyclic martian satellite hypothesis of \citet{hes17} has Phobos (and previous-generations of inner satellites) migrating outward from the edge of the ring at $3.2~R_M$, and then reversing migration before ever reaching the 2:1 resonance with Deimos, consistent with the observed low eccentricity of Deimos.

The inclination of Deimos presents a different challenge, as the satellite's almost 2-degree tilt is quite remarkable. High orbital inclinations in otherwise dynamically cold satellite systems are often explained by convergent migration of two satellites into a second-order mean motion resonance (MMR), resulting in a capture into an inclination type resonance \citep{tw2, mal90}. Deimos is not expected to migrate significantly due to tides \citep{yod82}, and should have migrated only a very small amount due to radiation force \citep[chiefly the Binary YORP effect;][]{cuk05}, even if we assume that Deimos's surface shape was unchanged over several Gyr. Therefore, we would need Phobos, or a more massive precursor inner satellite of Mars from an earlier cycle, to migrate outward through a second order resonance with Deimos. The innermost eligible resonance is the 3:1 MMR with Deimos at about 3.3 $R_M$, which could have plausibly been crossed by an inner satellites of Mars (including Phobos) migrating outward due to torques from the martian ring \citep{hes17}. In the next Section, we present numerical simulations of this resonance, and show that this process produces an exceptionally good match for the present eccentricity and inclinations of Deimos. In further sections we will briefly present the analysis of the dynamical mechanism involved, and discuss implications for the history of the martian satellite system. 

\section{Numerical Simulations} \label{sec:numerical}

For integrations of the 3:1 MMR between a hypothetical past martian inner satellite and Deimos we use the numerical integrator {\sc simpl}, which was extensively tested by \citet{cuk16}. {\sc simpl} is a N-body mixed-variable symplectic integrator that integrates the full equations of motion. We refer the reader to \citet{cuk16} for the details of the integrator, and we will here simply specify the parameters used in our simulations. We included the second-order oblateness moment of Mars $J_2$ (assumed to be same as current), the Sun (assuming a Keplerian orbit for Mars) and two satellites as point-masses. The two satellites in the model include Deimos, with its current mass, and an inner satellite, which we varied between $1-200\times$ Phobos mass. We also varied the obliquity of Mars and the eccentricity of the martian heliocentric orbit. We assumed that both satellites had very low eccentricities and inclinations at the time of the 3:1 resonance crossing (see Section \ref{sec:discussion}). We included both the tidal acceleration (which makes the inner satellite migrate inward), and a larger artificial tangential acceleration on the inner satellite that approximates ring torques, resulting in net outward migration. We otherwise ignored the ring's gravity, as a 100 $M_P$ ring would increase the martian obliquity moment $J_2$ by only about 0.1\%. We used a timestep of $4 \times 10^{-5}$~yr (0.35~h) in all simulations shown here.

\begin{figure*}
\epsscale{.6}
\plotone{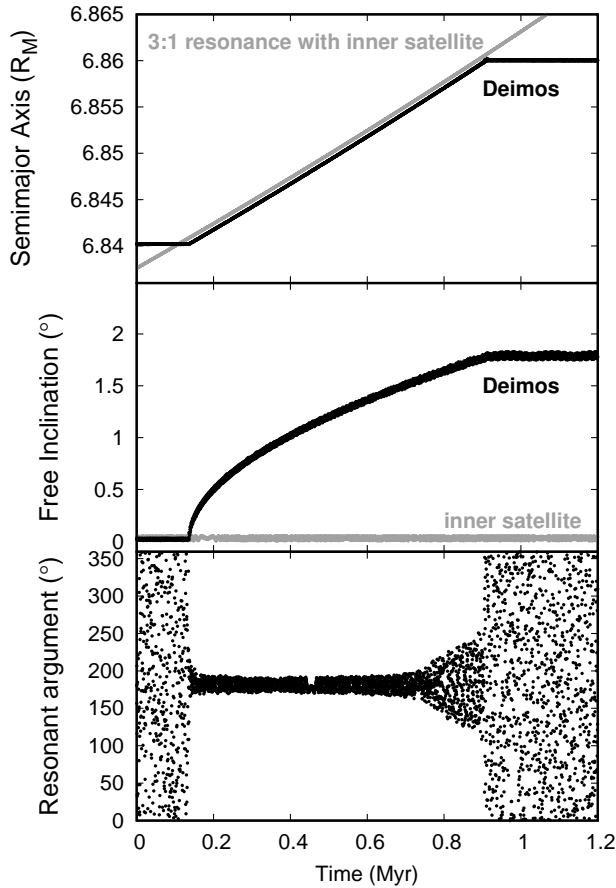}
\caption{Simulation of 3:1 MMR between an 18 Phobos-mass inner satellite of Mars and Deimos. The inner satellite is assumed to be migrating outward due to ring torques at a net rate that is 16\% of its inward tidal migration rate. In the top panel, the gray line shows the semimajor axis at which a satellite would be in exact 1:3 commensurability with the inner satellite. The bottom panel shows the resonant argument $3 \lambda - \lambda_I -\Omega_f -\Omega_L$ (see text). The resonance breaks due to libration growth caused by a secondary resonance starting at 0.7~Myr.\label{fig1}}
\end{figure*}

We find that capture into an inclination-type subresonance of 3:1 MMR is a common outcome for a range of masses, whenever the net orbital migration rate of the inner satellite is slow enough. The resonance capture is not indefinite, but the resonance breaks at Deimos inclinations between zero and three degrees, with Deimos's final inclination varying both stochastically and systematically. Figure \ref{fig1} shows the evolution of the satellites' semimajor axes and inclinations during the resonance capture for an inner satellite 18 times more massive than Phobos. The inclination of the inner satellite is not affected by this resonance, therefore implying that we are looking at $i^2$ type subresonance, in which only the inclination of Deimos is affected \citep[][Section 8.8.2]{md99}. However, we find that the relevant resonant argument is $3 \lambda - \lambda_I -\Omega_f -\Omega_L$ (Fig. \ref{fig1}, bottom panel), where $\lambda$ and $\Omega_f$ are the mean longitude and longitude of the node of Deimos, and $\lambda_I$ is the longitude of the inner satellite. The resonant argument also depends on the orientation of Deimos's Laplace plane $\Omega_L$ (measured with respect to the martian equator), which is unusual and will be discussed in the next Section.

\begin{figure*}
\epsscale{.8}
\plotone{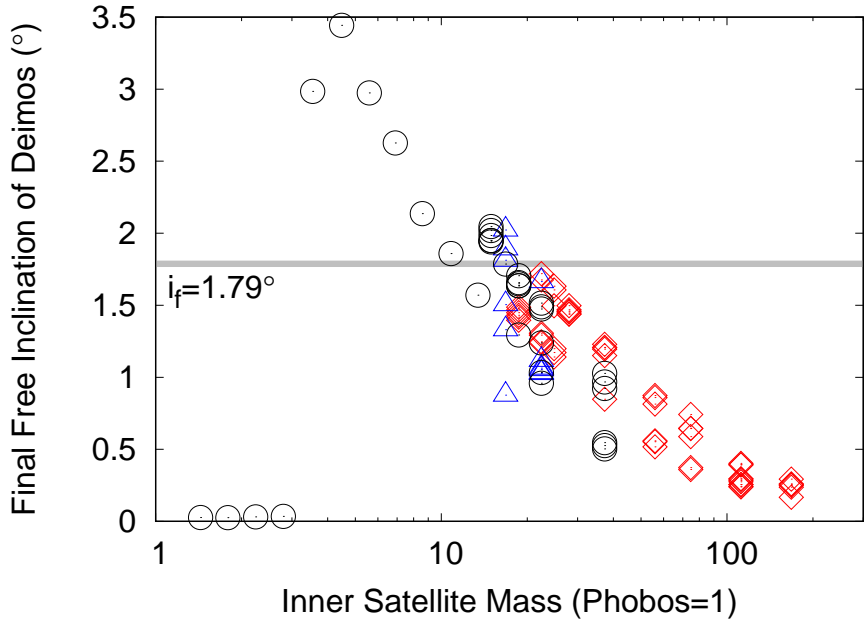}
\caption{Dependence of the final free inclination of Deimos on the mass of the inner satellite (expressed in Phobos masses $M_P$). Black circles and blue triangles plot simulations using martian obliquity $\epsilon=40^{\circ}$, while the red diamonds show runs with $\epsilon=25^{\circ}$. Blue triangles plot integrations using the martian heliocentric eccentricity $e_M=0.04$, all other cases have $e_M=0.09$. The horizontal gray line indicates the current free inclination of Deimos. Very low inclinations on the far left-hand side indicate that Deimos was not captured in the resonance in those simulations.\label{fig2}}
\end{figure*}

The dependence of the final inclination of Deimos after the resonance passage on the mass of the inner satellite is plotted in Fig. \ref{fig2}. Each point plots the end-state of a numerical simulation. For many inner satellite mass values, we ran sets of simulations that differed only in the separation of the system from the resonance, enabling us to survey the inherent stochasticity of the process. Red diamonds plot simulations that used a martian obliquity of $\epsilon=25^{\circ}$, comparable with the present state, while black circles plot simulations with martian obliquity of $\epsilon=40^{\circ}$, closer to the long-term average \citep{las04}. While those simulations used the martian heliocentric obliquity of $e_M=0.09$, blue triangles plot simulations with $e_M=0.04$ (with $\epsilon=40^{\circ}$); we find no systematic dependence of the resonant outcomes on martian eccentricity. Plotted simulations generally used net outward migration rates that were about $16\%$ of the (inward) tidal evolution rate. Fig. \ref{fig2} shows that there is an increase of final inclination of Deimos with decreasing mass of the inner satellite in the 3-200 Phobos mass ($M_P$) range. Less massive inner satellites do not capture Deimos into resonance at this relative migration rate, and Deimos is left with a very low final inclination.  

The simulations plotted in Fig. \ref{fig2} visibly cluster along discrete curves, and these curves are offset for simulation sets that used different obliquities of Mars. We find that this pattern is caused by secondary resonances, which increase the libration amplitude of the $i_f i_L$resonance, eventually leading to breaking of the resonance lock. A secondary resonance is visible in the bottom panel of Fig. \ref{fig1}, starting at 0.7~Myr into the simulation. These secondary resonances are very similar to those found by \citet{mal90} in the context of the Miranda-Umbriel 3:1 inclination-type MMR, and their origin will be discussed in the next Section. 

\section{Dynamics of the Resonance}\label{sec:resonance}

The planes of all satellite orbits precess around their respective Laplace planes, with the orientation of the Laplace plane being determined by the sum of perturbations on the orbit. Close-in satellites like Phobos are chiefly perturbed by the planet's oblateness, and their Laplace plane is very close to the planet's equator. Distant satellites like the Moon are overwhelmingly perturbed by the Sun, so their Laplace plane is almost identical to the plane of the planet's heliocentric orbit. Deimos is intermediate between these two regimes, as its Laplace plane is not exactly in the plane of the martian equator, but is inclined by about a degree toward Mars's heliocentric orbit. Starting with the gravitational potential derived in \citet{tam13} \citep[see also ][]{tre09}, we can derive Deimos's Laplace plane tilt by requiring the terms containing $\cos{\Omega}$ in the disturbing function to be zero (this is the requirement for uniform nodal precession): 
\begin{equation}
\label{laplace}
\sin{i_L}={M_{Sun} a^5 \over 2 J_2 M_M a^3_M R^2} \sin{\epsilon} \cos{\epsilon}
\end{equation}
where $M_{Sun}$ is the solar mass, and $M$, $a_M$, $R$ and $J_2$ are respectively the mass, semimajor axis, radius, and oblateness moment of Mars, and $a$ is Deimos's semimajor axis. We also assumed circular orbits and $i_L << \epsilon$. Eq. \ref{laplace} shows that the maximum $i_L$ happens at $\epsilon=45^{\circ}$, when $i_L \simeq 1.15^{\circ}$. 

The inclination-type 3:1 MMR we are interested in has (apart from terms with no angular dependence) a resonant term in the disturbing function of the type 
\begin{equation}
\label{res}
R_{res} = \sin^2{i} f_{57} \cos(3 \lambda - \lambda_I - 2 \Omega)
\end{equation}
where $f_{57}$ is a function of masses and semimajor axes \citep[][Table B.8]{md99}. These terms can be derived using the geometry of the interaction between the inner satellite and Deimos, and this derivation assumes that the inclinations and nodes of both orbits are measured relative to the same reference plane. This last assumption is incorrect, as the Laplace plane of Deimos is tilted by a non-trivial amount. In order to obtain an accurate description of mutual positions and gravitational interactions, we will take the martian equator as the reference plane, and recognize that $\sin{i}\sin{\Omega}=\sin{i_f}\sin{\Omega_f} + \sin{i_L}\sin{\Omega_L}$ and $\sin{i}\cos{\Omega}=\sin{i_f}\cos{\Omega_f} + \sin{i_L}\cos{\Omega_L}$, where the subscript $f$ now refers to the ``free'' inclination and node of Deimos \citep[][Section 7.4]{md99}. We can then transform the resonant argument in Eq. \ref{res} so that we separate factors of $\sin{i}\cos{\Omega}$ and $\sin{i}\sin{\Omega}$, express them in terms of free and forced components, and then reconstitute the new resonant terms. It can be shown that the original term produces three new resonant arguments, and the one we are interested in is    
\begin{equation}
\label{res2}
R_{res}' = 2 \sin{i_L}\sin{i_f} f_{57} \cos(3 \lambda - \lambda_I - \Omega_f - \Omega_L).
\end{equation}
Therefore, due to the unusual Laplace plane of Deimos, we have a resonance that is derived from a pure inclination ($i^2$-type) subresonance of the 3:1 MMR, but in many ways behaves as a mixed $i i'$-type resonance. Physically this makes sense, as the inner satellite is a perturber on an inclined orbit, despite this tilt being due to orientation of Deimos's Laplace plane, and not the inner satellite's own inclination.  

The above description of the resonance enables us to understand the secondary resonances that lead to Deimos exiting the resonance, as seen in Fig. \ref{fig1}. The closest neighbor of our $i_f i_L$ resonance is the pure free-inclination subresonance $i_f^2$ with the resonant argument $3 \lambda - \lambda_I - 2 \Omega_f$, which is circulating with Deimos's orbital precession period of about 54~yr while the resonant argument of $i_f i_L$ resonance is librating. The fact that our simulations do not include Mars's axial and orbital precession is not important, as in reality $\dot{\Omega}_L/\dot{\Omega}_f < 10^{-3}$. In our simulations we find that the system can be captured into a secondary resonance when the libration frequency in the $i_f i_L$ resonance is a simple multiple of the circulation frequency of the $i_f^2$ argument. The libration period becomes shorter with increasing $i_f$, and longer for larger amplitudes of libration of the resonant argument. Therefore, in order to maintain the secondary resonance as the free inclination grows, the amplitude of libration of the $i_f i_L$ resonance has to grow over time. This widening of the libration amplitude is visible in Fig. \ref{fig1} starting at 0.7~Myr, and eventually leads to breaking of the resonant lock. 

In Fig. \ref{fig1}, the libration frequency at which the secondary resonance happens has a period of about 108~years, making this a 2:1 secondary resonance. In Fig. \ref{fig2}, we can see the signature of the 2:1 secondary resonance as the topmost lines our simulations cluster around. As the plot suggests, the same secondary resonance happens at higher free inclinations for smaller masses, as librations in $i_f i_L$ resonance slow down with decreasing mass. The 2:1 secondary resonance also tends to happen at higher free inclination for smaller obliquities, as the line corresponding to the 2:1 secondary resonance for red diamonds (integrated at $\epsilon=25^{\circ}$) is higher than that for black circles ($\epsilon=40^{\circ}$). The line below that of the 2:1 secondary resonance corresponds to the 3:1 secondary resonance (i.e. libration periods of about 162~yr), and a few points hint at higher-order secondary resonances. The distribution of probabilities for capture into each of these resonances vary with the inner satellite mass, and while for some inner satellite masses (like $18 M_P$ used in Fig. \ref{fig1}) we get one secondary resonance to be dominant, for the inner satellite masses there is a diversity of outcomes. 

\section{Discussion and Conclusions} \label{sec:discussion}
 
\begin{figure*}
\epsscale{.75}
\plotone{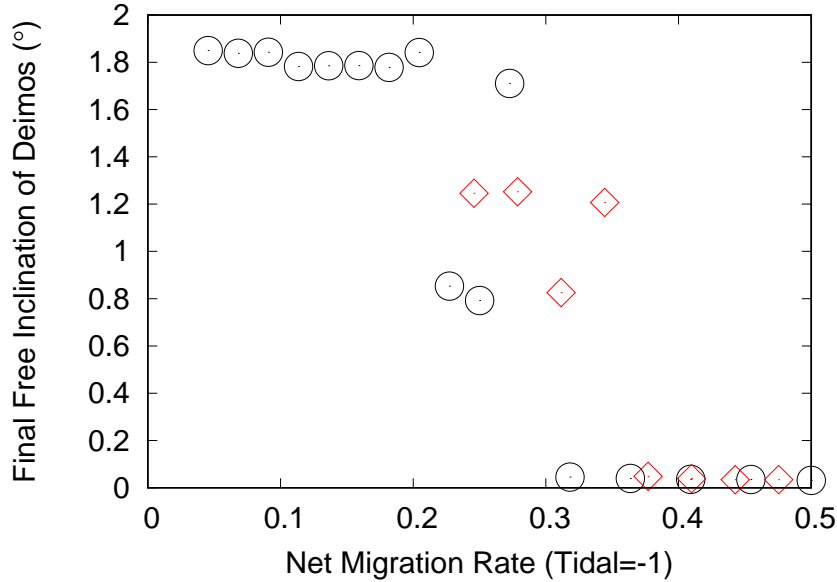}
\caption{Effects of the inner satellite's net migration rate (normalized to tidal) on the final free inclination of Deimos. One set of simulations uses inner satellite mass of $M_I=18 M_P$ and the martian obliquity of $\epsilon=40^{\circ}$ (black circles), while the other set has $M_I=25 M_P$ and $\epsilon=25^{\circ}$. Tidal migration rate has been calculated using tidal quality factor $Q=80$ and tidal Love number $k_2=0.14$ for Mars. High final inclinations of Deimos indicate resonance capture, while low values reflect resonance crossing without capture. While the transition between resonance ``jump" and capture is dynamically significant, the variation of final inclinations on the left-hand side may be stochastic and reflect probabilistic capture into different secondary resonances, rather than sensitive dependence on the relative migration rate. \label{fig3}}
\end{figure*}

While we have demonstrated in Figs. \ref{fig1} and \ref{fig2} that capture into an inclination-type 3:1 MMR between an inner satellite and Deimos is plausible, resonance capture is known to sensitively depend on the rate of orbital evolution. To examine this dependence, we ran two sets of simulations with varying net migration rates, one featuring a $M_I=18 M_P$ inner satellite and martian obliquity $\epsilon=40^{\circ}$ (same as in Fig. \ref{fig1}), and the other with a $M_I=25 M_P$ inner satellite and martian obliquity of $\epsilon=25^{\circ}$. The resulting free inclinations of Deimos generated in these simulations are plotted in Fig. \ref{fig3}. The maximum net migration rate at which resonance capture still occurs is at about 30-35\% of the tidal migration rate (itself proportional to the inner satellite's mass). The relative critical migration rate is larger for the more massive inner satellite, implying a dependence on the inner satellite mass (the difference in martian obliquity in Fig. \ref{fig3} is expected to have the opposite effect, favoring the $M_I=18 M_P$, $\epsilon=40^{\circ}$ simulations). One direct implication is that the boundary between capture and non-capture into resonance seen on the left-hand side of Fig. \ref{fig2} is not an intrinsic property of the inner satellite mass, but depends on the migration rate. We ran an additional simulation and confirmed that an inner satellite with $M_I=3 M_P$ can capture Deimos into the $i_f i_L$ resonance, if the net migration is about 8\% of the tidal evolution rate. Therefore, resonance capture is still possible at very low masses of the inner satellite, but increasingly closely matched ring-torques and tidal acceleration are necessary in order for it to happen.

More broadly, what is the likelihood of this scenario in which the inner satellite migrates outward through the 3:1 resonance with Deimos at $3.3 R_M$ at a fraction of the tidal evolution rate? We would expect slow evolution to happen just before the ring torques become too weak to overcome the inward tidal acceleration, and the orbital migration reverses. \citet{hes17} envision inner satellite formation from the outer edge of the ring at the Fluid Roche limit, which is $\sim3.1-3.4 R_M$ for material with mass density in the range of that of Phobos and Deimos ($\sim1.5-1.9~\mathrm{g\ cm^{-3}}$). Interestingly, \citet{hu20} recently suggested that the overall shape of Phobos is the best match to tidal forces at $\sim3.3 R_M$, rather than its present distance, suggesting formation or a prolonged period of residence at that distance, consistent with this model. Our model would favor a formation of the inner satellite at $<3.3 R_M$, followed by some amount of outward orbital evolution from ring torques through the 3:1 MMR. While the exact numbers are model-dependent, \citet{hes17} do expect the last three generations of the inner satellites (including Phobos) to only migrate out to about that distance.  As the ring would be dissipating, the ring torques would become weaker over time, and a very slow outward migration through the 3:1 MMR with Deimos at $3.3 R_M$ is certainly plausible in the context of the model of ring-satellite cycling. We also note that tidal dissipation within Mars may have been weaker in the distant past \citet{sam19}, relaxing some of the constraints on the near-canceling of ring and tidal torques.

\begin{figure*}
\epsscale{.8}
\plotone{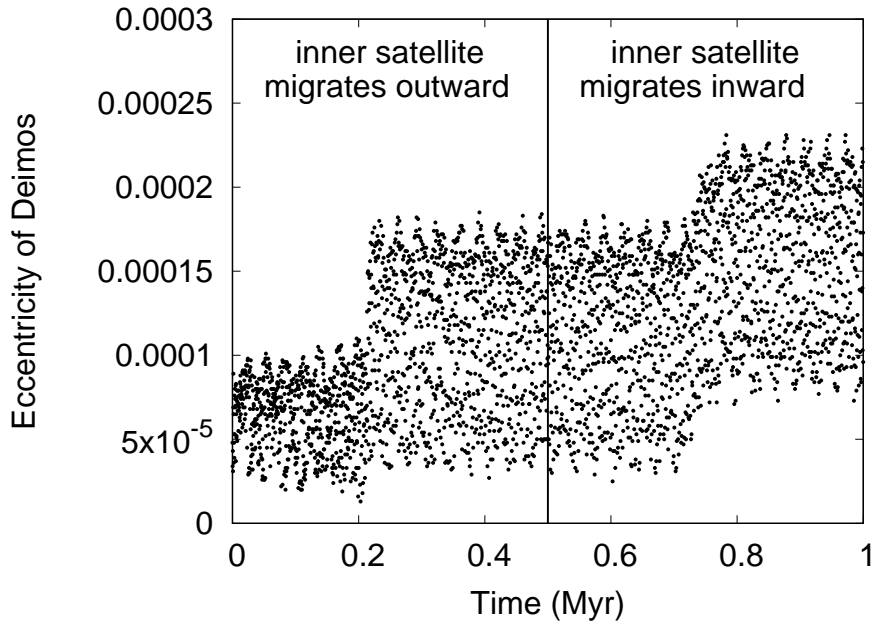}
\caption{Two excerpts from the same integration (as shown in Fig. \ref{fig1}) of a $18 M_P$ inner satellite crossing the 3:1 MMR with Deimos at 16\% of its tidal migration rate (martian obliquity of $40^{\circ}$). The left-hand side shows the crossing of the $e^2$ sub-resonance of the 3:1 MMR that closely follows the capture into the $i_f i_L$ subresonance shown in Fig. \ref{fig1}. Subsequently, we reduced the outward torque and the inner satellite migrated back through the 3:1 MMR with Deimos, with the second crossing shown on the right-hand side. The modern-day eccentricity of Deimos is $e=2.7 \times 10^{-4}$ \citep{jac14}.\label{fig4}}
\end{figure*}

Apart from the inclination, the 3:1 MMR between the inner satellite and Deimos also affects Deimos's eccentricity. Since there is no equivalent of Laplace plane tilt for eccentricity, the strongest eccentricity subresonance of 3:1 MMR is the $e^2$ \citep[][ Section 8.8.2]{md99}. We do not see capture in that resonance in any of our simulations, but instead Deimos experiences a kick to its eccentricity. Figure \ref{fig4} shows the eccentricity of Deimos at the moment of the 3:1 MMR crossing in the same simulation as shown in Fig. \ref{fig1}, with the additional integration of the reverse passage of the inner satellite through the same resonance during its later inward migration. In general, the addition of eccentricity jumps generated in the two passages is random, so sometimes they add up and sometimes they cancel out. We conclude that the excitation of Deimos's eccentricity by resonance passages by a "Cycle 3" satellite \citep[in the terminology of ][]{hes17} is comparable and consistent with Deimos's current $e=2.4 \times 10^{-4}$. 

Fig. \ref{fig2} suggests that the best match to Deimos's inclination is obtained by a slow outward migration of a satellite with a mass of about $M_I=20 M_P$ through the 3:1 MMR with Deimos, assuming that both satellites have had circular and planar orbits at that time. How does this situation fit into the broader understanding of the history of the martian system? This mass of the inner satellite is close to that predicted for the ``Cycle 3" satellite by \citet{hes17} (numbering goes backward in time; Phobos itself is Cycle 1). This is also expected to be the first generation of past martian inner satellites to contain only one satellite, which is consistent with undisturbed evolution through this relatively weak resonance with Deimos. As Deimos would have its current inclination by the time the Cycle 2 satellite and Phobos encounter the 3:1 resonance with Deimos, resonance capture is excluded, and only small kicks to $e$ and $i$ or Deimos are expected. 

As the inner satellite would have recently formed from the ring, we would expect its orbit to be circular and planar. However, Deimos may have experienced resonances with past generations of inner satellites, making its low inclination at the time problematic (note that this is a problem for the eccentricity of Deimos in any scenario involving migrating massive inner satellites). While more modeling work is needed to understand this, we speculate that multiple inner satellites of Cycle 4 \citep[as per][]{hes17} excited the eccentricity and/or inclination of Deimos enough that outside bombardment (heliocentric or planetocentric) triggered a runaway cascade of sesquinary impacts which damped the eccentricity and inclination of Deimos, after which re-accretion was possible \citet{nay16}. According to \citet{hes17}, re-accretion during the Cycle 4 would make Deimos about 3.5-4 Gyr old, which would be broadly consistent with much of its surface craters being heliocentric \citep{tho80}, although we would also expect a significant sesquinary component \citep{nay16}. 

To conclude, we find that the present orbital inclination and eccentricity of Deimos strongly suggest a past 3:1 mean-motion resonance with an outward migrating past inner satellite of Mars about 20 times more massive than Phobos. Since the 3:1 MMR with Deimos is well within the synchronous orbit, this outward migration requires dynamical interaction with a substantial ring of Mars at the time (3-4 Gyr ago). Our results strongly support the cyclic martian satellite hypothesis of \citet{hes17}. Broader implications include a much larger mass of circumplanetary material around early Mars than is currently in Phobos and Deimos, and a relatively young age for Phobos, requiring most of the craters on its surface to be made by planetocentric impactors.  

\acknowledgments

M\'C, DAM, and CW are supported by NASA Emerging Worlds Program award 80NSSC19K0512. An anonymous reviewer helped to significantly improve the manuscript. We wish to thank all the people who are working during the COVID-19 pandemic to keep the rest of us safe.

\bibliography{deimos_refs}{}
\bibliographystyle{aasjournal}

\end{document}